\documentclass[11pt]{article}
\newcommand{\comm}[1]{}

\usepackage{amssymb}%,psfig}
\usepackage{amsthm}
\usepackage[centertags]{amsmath}
\usepackage{amsfonts}
\usepackage{amssymb}
\usepackage{amsthm}
\usepackage{epsfig}
\usepackage{setspace}
\usepackage{ae} % Umlauts
\usepackage{eucal} % I think standard calligraphics are horrible
\def\citet{\cite}

\newtheorem{theorem}{Theorem}%[section]
\newtheorem{lemma}{Lemma}%[section]
\newtheorem{corollary}{Corollary}%[section]

\newtheorem{definition}{Definition}%[section]
\newtheorem{example}{Example}%[section]
\newcounter{thanksnum}
\def\thanksnumber#1
{\setcounter{thanksnum}{\value{footnote}}\setcounter{footnote}{#1}%
                     \addtocounter{footnote}{-1}\footnotemark
                     \setcounter{footnote}{\value{thanksnum}}}
\def\newtheoremz#1{\@ifnextchar[{\@othmz{#1}}{\@nthmz{#1}}}

\def\@nthmz#1#2{%
\@ifnextchar[{\@xnthmz{#1}{#2}}{\@ynthmz{#1}{#2}}}

\def\@xnthmz#1#2[#3]{\expandafter\@ifdefinable\csname #1\endcsname
{\@definecounter{#1}\@addtoreset{#1}{#3}%
\expandafter\xdef\csname the#1\endcsname{\expandafter\noexpand
  \csname the#3\endcsname \@thmcountersepz \@thmcounterz{#1}}%
\global\@namedef{#1}{\@thmz{#1}{#2}}\global\@namedef{end#1}{\@endtheoremz}}}

\def\@ynthmz#1#2{\expandafter\@ifdefinable\csname #1\endcsname
{\@definecounter{#1}%
\expandafter\xdef\csname the#1\endcsname{\@thmcounterz{#1}}%
\global\@namedef{#1}{\@thm{#1}{#2}}\global\@namedef{end#1}{\@endtheoremz}}}

\def\@othmz#1[#2]#3{\expandafter\@ifdefinable\csname #1\endcsname
  {\global\@namedef{the#1}{\@nameuse{the#2}}%
\global\@namedef{#1}{\@thmz{#2}{#3}}%
\global\@namedef{end#1}{\@endtheoremz}}}

\def\@thmz#1#2{\refstepcounter
    {#1}\@ifnextchar[{\@ythmz{#1}{#2}}{\@xthmz{#1}{#2}}}

\def\@xthmz#1#2{\@begintheoremz{#2}{\csname the#1\endcsname}\ignorespaces}
\def\@ythmz#1#2[#3]{\@opargbegintheoremz{#2}{\csname
       the#1\endcsname}{#3}\ignorespaces}

\def\@thmcounterz#1{\noexpand\arabic{#1}}
\def\@thmcountersepz{.}
\def\@begintheoremz#1#2{ \trivlist \item[\hskip \labelsep{\bf #1\ #2}]}
\def\@opargbegintheoremz#1#2#3{ \trivlist
      \item[\hskip \labelsep{\bf #1\ #2\ (#3)}]}
\def\@endtheoremz{\endtrivlist}

\def\e{\varepsilon}

\def\d{\delta}

\def\O{\Omega}

\def\F{{\cal F}}
\def\w{\widehat}

\def\R{{\bf R}}
\def\E{{\bf E}}
\def\P{{\bf P}}

\def\T{{\cal T}}

\def\b{\beta}
\def\s{\delta}
\def\g{\gamma}

\def\t{\theta}
\def\oo{\bar}
\def\s{\sigma}
\def\G{\Gamma}
\def\GG{{\cal G}}

\def\A{{\cal A}}
\def\M{{\cal M}}

\newcommand{\be}{\begin{equation}}
\newcommand{\ee}{\end{equation}}
\newcommand{\bd}{\begin{displaymath}}
\newcommand{\ed}{\end{displaymath}}
\newcommand{\ba}{\begin{array}{ll}}
\newcommand{\ea}{\end{array}}
\newcommand{\baa}{\begin{eqnarray}}
\newcommand{\eaa}{\end{eqnarray}}
\newcommand{\baaa}{\begin{eqnarray*}}
\newcommand{\eaaa}{\end{eqnarray*}}   \font\sm=cmr10
\def\k{\kappa}

\def\Ts{s}
\def\DD{{\cal D}}
\def\Td{{\cal T}_\d}
\def\TT{\Theta}
\def\AA{{\rm A}}
\title{On the no-arbitrage market and continuity in the  Hurst parameter }
\author{
Nikolai Dokuchaev\\
 {\sm Department of Mathematics \& Statistics, Curtin
University}\\
{\sm  email: N.Dokuchaev@curtin.edu.au } }
\begin{document}
\maketitle
\begin{abstract}
We  consider a  market with fractional Brownian motion  with stochastic integrals generated by the Riemann  sums. We found that this market is arbitrage free if admissible strategies that are using observations with an arbitrarily small delay.
Moreover, we found that this approach  eliminates the  discontinuity with respect to
the Hurst parameter $H$ at  $H=1/2$ of the expectations of stochastic integrals.
\par
 {\bf Key words}: market models, portfolio selection,
 fractional  Brownian motion, arbitrage, arbitrage-free market.
\par
{\bf JEL classification}: C52, % - Model Evaluation, Validation, and Selection
C53, % - Forecasting Models; Simulation Methods
%C54 - Quantitative Policy Modeling
%C58 - Financial Econometrics
%C61 - Optimization Techniques; Programming Models; Dynamic Analysis
G11% - Portfolio Choice; Investment Decisions
\par
{\bf Mathematics Subject Classification (2010)}:
 	91G70, %	Statistical methods, econometrics
 % 91B24, %   	Price theory and market structure
 60G22,   %	Fractional processes, including fractional Brownian motion
 91G10    %	Portfolio theory
\end{abstract}
%see(?)http://www.artofproblemsolving.com/LaTeX/AoPS_L_GuidePack.php
 \section{Introduction}
In this short note, we readdress the problem of the presence of arbitrage opportunities for the market models based
on fractional Brownian motion with the Hurst parameter $H\in (1/2,1)$. Statistical properties of these models make them important for financial applications; however, the presence  of arbitrage represents a certain obstacle from the theoretical point of view. This problem was intensively studied; see, e.g., \cite{BKiev,BPS,B11,Bj,C,G,Ma,R,Shi,Sal}. As can be seen from Example \ref{ex1} below, there is a discontinuity with respect to $H\to 1/2+0$   at the point $H=1/2$ of the wealth process   for some portfolio strategies.  The market where $H=1/2$ is arbitrage free, and the market with $H\in(1/2,1)$ allows arbitrage.
One of possible  some solutions  of this problem  is to use different constructions  of stochastic integral that are not based on Riemann sums
such as Wick integral (see \cite{BKiev,Bj}).  Another approach is to include proportional transaction costs in the model \cite{G,BPS}.
In addition,  it was suggested in \cite{C} that additional restrictions on the admissible strategies also can remove arbitrage. It was shown
in Theorem 4.3 \cite{C} that arbitrage cannot be achieved in the class of piecewise constant strategies with a minimal amount of
time between two consecutive transactions. The restrictions on the times between transactions were relaxed in \cite{B11}, Theorem 3.21.

 We suggest one more alternative class of strategies allowing to exclude arbitrage for a
 a  market based on a fractional Brownian motion with $H\ge 1/2$ with stochastic integrals generated by the Riemann  sums.
 We suggest to use admissible strategies that are not necessary piecewise constant and that
 they are constructed  using current observations processed with an arbitrarily small time delay.
   It can be noted that this is a natural restrictions on the class of the portfolio strategies;  in practice, certain delay in information transfer and execution is inevitable for practical implementation of a portfolio  strategy.
\par
We found that a simple  Bachelier type market of with these strategies is arbitrage free (Theorem \ref{ThNA}); this result is similar to
 to the results for piecewise constant strategies from  Theorem 4.3 \cite{C} and Theorem 3.21 \cite{B11}.
 \par
The most interesting result of this paper is that it appears  the  discontinuity with respect to $H$ at  $H=1/2$ of the expectations of stochastic integrals vanishes for our class strategies (Lemma \ref{ThE} and Theorem \ref{ThNA}(ii)).
\par
The proofs are based on a useful representation for
the fractional Brownian motion $B_H(t)$  with the zero mean.  We found that the increment  of $B_H(t)$ can be  represented as the sum of a two independent Gaussian
processes  one of  which is smooth in the sense that it is differentiable  in mean square sense,
with the derivative that is square integrable on the finite time intervals. Similarly to the drift part of the diffusion processes, expectations of the
integrals by
this process are non-zero for the processes adapted to it.
 This process can be considered as an analog of the drift. It has to be noted that  the term "drift" is usually applied to
 $\mu$ presented  for the process $\mu t+B_H(t)$; see
\cite{CNS},\cite{Mu},\cite{Es}, where  and estimation of $\mu$ was studied.
In  \cite{Mu}, the term "drift" was also used for a representation for $B_H$ after linear integral transformation and random time change via a standard Brownian motion process with constant in time drift.
 Our representation is for $B_H$ itself, i.e, without an integral transformation.
\section{The  main result}
\subsection*{The fractional Brownian motion and stochastic integration}
 We are
given a standard probability space $(\Omega,\F,\P)$, where $\Omega$ is
a set of elementary events, $\F$ is a complete $\s$-algebra of
events, and $\P$ is a probability measure.

 We  assume that $\{B_H(t)\}_{t\in\R}$ is a fractional Brownian motion such that $B_H(0)=0$
 with the Hurst parameter $H\in  (1/2,1)$ defined as described in \cite{Ma,GN}
 such that
 \baa
 B_H(t)-B_H(s)=c_H\int_s^t(t-q)^{H-1/2}dB(q)+c_H  \int_{-\infty}^s\left[(t-q)^{H-1/2}-(s-q)^{H-1/2}\right]dB(q),
 \label{BBH}
 \eaa
 where $t>s>0$,
 $c_H=\sqrt{2H\G (3/2-H)/[\G(1/2+H)\G(2-2H)]}$ and  $\G$ is the gamma function. Here
$\{B(t)\}_{t\in\R}$ is  standard Brownian motion such that $B(0)=0$.
\par
Let $\{\GG_t\}$ be the filtration generated by the process $B(t)$.

For a given $T>s$, let $\AA_0(s,T)$ be the set of all  processes  $\g(t)$, $t\in[s,T]$,
that are progressively
measurable with respect to the filtration $\{\GG_t\}$ and such that $\E\int_0^T\g(t)^2dt<+\infty$.

Let $\AA_\e[s,T]$  be the set of all  $\g\in\AA_0[s,T]$
such that there exists an integer $n>0$ and a set of non-random times
$\T=\{T_k\}_{k=1}^n\subset [s,T]$, where $n>0$ is an integer, $T_0=s$, $T_n=T$,
and $T_{k+1}-T_k\ge \e$, such that $\g(t)$ is $\GG_{T_k}$-measurable for $t\in[T_k,T_{k+1})$.
In particular, this set includes all $\g\in\AA_0[s,T]$ such that $\g(t)$ is $\GG_{t-\e}$-measurable for all $t\in[s,T]$.
Let $\AA^d[s,T]=\cup_{\e>0}\AA_\e[s,T]$.

Let $\w\AA_\e$ the set of all  $\g\in\A_0$ such that $\g(t)$ is $\GG_{t-\e}$-adapted, and let $\w\AA^d=\cup_{\e>0}\w\A_\e$.

\begin{lemma}\label{lemmaI} For any $\g\in\AA^d[s,T]$, the integral $\int_s^T\g(t)dB_H(t)$ converge as a sequence of the corresponding
Riemann sums  in $L_1(\Omega,\GG_{T},\P)$.
\end{lemma}
The following lemma establishes continuity in $L_1(\Omega,\GG_{T},\P)$  of the stochastic integrals with respect to
the Hurst parameter $H$ at  $H=1/2$.
 \begin{lemma}\label{ThE}
For any $\g\in\w\AA^d$, \baa
\E\left|\int_s^T\g(t)dB_H(t)-\int_s^T\g(t)dB(t)\right|\to 0 \quad\hbox{as}\quad H\to 1/2+0.
\label{lim}\eaa \end{lemma}
\par
This continuity does not take place for some $\g$, as can be seen from the following example
\cite{Shi}.
\begin{example}\label{ex0}
For any $H\in(1/2,1)$ and $T>s$,
\baaa
(B_H(T)-B_H(s))^2=2\int_0^T(B_H(t)-B_H(s))dB_H(t).
\label{sqB}\eaaa
The integral converges  as a sequence of the corresponding
Riemann sums.
\end{example}
A  consequence of  Example  \ref{ex0} is that
\baaa
\E\int_s^T\g(t)dB_H(t)\nrightarrow 0=\E\int_s^T\g(t)dB(t) \quad\hbox{as}\quad H\to 1/2+0.
\eaaa
Therefore, the stochastic integrals by $dB_H$
depends discontinuously in $H\to 1/2+0$ for some $g\in\AA_0[s,T]$.

By Lemma \ref{ThE}, it also follows from Example  \ref{ex0}
that,  for $\g_\e(t)=\E\{g(t)|\GG_{t-\e}\}$,
\baaa
\E\int_s^T\g_\e(t)dB_H(t)\nrightarrow \E\int_s^T\g(t)dB_H(t) \quad\hbox{as}\quad \e\to 0+.
\eaaa
Therefore, the stochastic integrals by $dB_H$
depends discontinuously in $\e$ with these $g_\e$.

\subsection*{The market model}
The rules for the operations of the agents on the market define the
class of admissible strategies  where the optimization problems have
to be solved.

Let $X(0)>0$ be the initial wealth at time $t=0$ and let $X(t)$ be
the wealth at time $t>0$.
\par
We assume that the wealth $X(t)$ at time $t\in[0,T]$ is
\begin{equation}
\label{X} X(t)=\b(t)b(t)+\g(t)S(t).
\end{equation}
Here $\b(t)$ is the quantity of the bond portfolio, $\g(t)$ is the
quantity of the stock  portfolio, $t\ge 0$. The pair $(\b(\cdot),
\g(\cdot))$ describes the state of the bond-stocks securities
portfolio at time $t$. Each of  these pairs is  called a strategy.

\par
 Let $\t\in(0,+\infty]$ be given; the case where $\t=+\infty$ is note excluded.

 Let $\{\F_t\}_{t\ge -\t}$
be a filtration such that $\F_t\subseteq \GG_t$ for all $t$.

\par  A pair $(\b(\cdot),\g(\cdot))$  is said to be an admissible
strategy if the processes $\b(t)$ and $\g(t)$ are progressively
measurable with respect to the filtration $\{\F_t\}$.

 In particular, the agents are not supposed to know the
future (i.e., the strategies have to be adapted to the flow of
current market information).

In addition, we require that  $$ \E\int_0^{T}\left[\b(t)^2+
\g(t)^2\right]dt<+\infty.$$
This restriction bounds the risk to be accepted and  pays the same role as exclusion of doubling strategies; see examples and
discussion  on doubling strategies in \cite{B11}.

\begin{definition}\label{defS}
\begin{enumerate}
\item
Let $\A_0$ be the set of all $\g$ that are progressively measurable  with respect to $\{\GG_t\}$
 such as described above.
\item
Let $\A_\e$ be the set of all  $\g\in\A_0$ such that there exists a finite set of non-random times $\T=\{T_k\}_{k=1}^n\subset [0,T]$, where $n>0$ is an integer, $T_0=0$, $T_n=T$,
and $T_{k+1}-T_k\ge \e$, such that  $\g(t)$ is $\GG_{T_k}$-measurable for $t\in[T_k,T_{k+1})$.
\item
Let $\A^d=\cup_{\e>0}\A_\e$.
\item
Let $\w\A_\e$ the set of all  $\g\in\A_0$ such that $\g(t)$ is $\GG_{t-\e}$-adapted.
\item
Let $\w\A^d=\cup_{\e>0}\w\A_\e$.
\end{enumerate}
\end{definition}

Note that $\w A_\e\subset \A_\e$ for any $\e>0$, and
the set $\A^d$  is wider than  the class of piecewise constant functions  considered in \cite{C}.

Suppose that, for some $\g\in\A_0$, the integral $\int_0^t\g(s)dS(s)$ converge as a sequence of the corresponding
Riemann sums. In this case, an  admissible
pair $(\b(\cdot),\g(\cdot))$  is said to be an admissible
self-financing strategy if
 $$
dX(t)=\b(t)db(t)+\g(t)dS(t)=\g(t)dS(t),
$$
meaning that
 $$
X(t)=X(0)+\int_0^t\b(s)db(s)+\int_0^t\g(s)dS(s)=\int_0^t\g(s)dS(s),
$$
 Under this condition, the process
$\g(t)$ alone defines the strategy.

By Lemma \ref{lemmaI}, for any $\g\in\A^d$, the integral $\int_0^T\g(s)dS(s)$ converge as a sequence of the corresponding
Riemann sums  in $L_1(\Omega,\GG_{T},\P)$.

Let $\A$ be a set of admissible $\g$  (we will consider  $\A=\A_0$, $\A=\A^d$,  or $\A=\w\A^d$).

For  $H\in [1/2,1)$, we denote by $\M_{H}(\A)$ the market model described above with  $\A$ as the set of admissible $\g$ .

\begin{definition}\label{defNA} We say that the market model $\M_{H}(\A)$ allows arbitrage if there exists a strategy $\g\in\A$ such that the
integral $\int_0^t\g(s)dS(s)$ converges  as a sequence of the corresponding
Riemann sums, and   $\P(X(T)\ge 0)=1$ and $\P(X(T)>0)>0$ for the corresponding self-financing strategy with the wealth $X(T)=\int_0^t\g(s)dS(s)$ at time $T$ with the initial wealth $X(0)=0$.
\end{definition}

It is known that the market model $\M_{1/2}(\A_0)$ does not allow  arbitrage.
On the other hand,  the market model $\M_{H}(\A_0)$ allows arbitrage for any $H\in (1/2,1)$.
This can be seen from
 the following version of Example \ref{ex0} \cite{Shi}.
\begin{example}\label{ex1}
For any $H\in(1/2,1)$,
\baa
X(T)=(S(T)-S(0))^2=2\int_0^T(S(t)-S(0))dS(t)=\int_0^T\g(t)dS(t),
\label{sq}\eaa
is the wealth for an admissible strategy with $\g(t)$ selected as $2(S(t)-S(0))$. In this case, the integral integral $\int_0^T\g(s)dS(s)$ converges  as a sequence of the corresponding
Riemann sums.
\end{example}

A consequence of  Example  \ref{ex1} is that the stochastic integrals by $dB_H$
depends discontinuously in $H\to 1/2+0$, since it is not true that
\baaa
\E\int_0^T\g(t)dB_H(t)\to 0=\E\int_0^T\g(t)dB(t) \quad\hbox{as}\quad H\to 1/2+0.
\eaaa
This is an undesired feature; small deviations of the evolution
law for $B_H$  cause large changes of the wealth for a strategy.
In addition, it implies that non-arbitrage model $\M_{1/2}(\A_0)$ and arbitrage allowing model $\M_{H}(\A_0)$ are
statistically indistinguishable  for the case where $H\approx 1/2$.
\begin{theorem}\label{ThNA}
\begin{enumerate}
\item
For any $H\in [1/2,1)$, the market model $\M_{H}(\A^d)$ is arbitrage-free.
\item
For any $\g\in\w\A^d$, \baaa
\E\left|\int_0^T\g(t)dB_H(t)-\int_0^T\g(t)dB(t)\right|\to 0 \quad\hbox{as}\quad H\to 1/2+0.
\label{lim2}\eaaa
\end{enumerate}
\end{theorem}
\section{Proofs}
Let $s\ge 0$ be fixed.
By (\ref{BBH}), we have that
\baaa
 B_H(t)-B_H(s)=W_H(t)+R_H(t),
 \eaaa
where
 \baaa
 W_H(t)=c_H\int_{s}^t(t-q)^{H-1/2}dB(q),\qquad
R_H(t)=c_H  \int_{-\infty}^{s}f(t,q)dB(q),
\label{BBHa}\eaaa
and where $f(t,q)=(t-q)^{H-1/2}-(s-q)^{H-1/2}$.

Let $s\ge 0$ and $T>s$ be fixed.
For $\tau\in [s,T]$ and  $g\in L_2(s,T)$, set \baaa
G_H(\tau,s,T,g)=c_H(H-1/2)\int_{\tau}^T(t-\tau)^{H-3/2}g(t)dt.
\eaaa
\begin{lemma}\label{ThM}
The processes $W_H(t)$ and $R_H(t)$, where $t>s$, are independent Gaussian $\{\GG_t\}$-adapted processes with zero mean and such that
 the following holds.
 \begin{enumerate}
 \item
 $W_H(t)$ is independent on $\GG_s$ for all $t>s$ and has an It\^o's differential in $t$ in the following sense:  for any $T>s$,
 there exists a function $h(\cdot,s,T)\in L_2(s,T)$ such that
 \baaa
 \int_{s}^T\g(t)dW_H(t)=\int_{s}^TG_H(\tau,s,T,\g)d B(\tau)
\eaaa
for any $\g\in L_2(\O,\GG_s,\P,L_2(s,T))$. 
\item
 $R_H(t)$ is $\GG_s$-measurable for all $t>s$ and differentiable in $t>s$ in mean square sense. More precisely,
 there exits a process   $\DD R_H$ such that
 \subitem(a)  $\DD R_H(t)$ is $\GG_s$-measurable for all $t>s$;
\subitem(b) for any $t>s$, \baaa \E \DD R_H(t)^2=c_H^2\frac{H-1/2}{2}(t-\Ts)^{2H-2},\qquad
\E\int_{s}^t \DD R_H(q)^2dq<+\infty;
\eaaa
\subitem(c)
for any $t>s$,  \baa
\lim_{\d\to 0} \E\left|\frac{R_H(t+\d)-R_H(t)}{\d} -\DD R_H(t)\right|=0.\label{deflim}
\eaa
\end{enumerate}
\end{lemma}
For $s\ge 0$, $T>s$, $\tau\in [s,T]$,  $g\in L_2(s,T)$, set \baa
G_H(\tau,s,T,g)=c_H(H-1/2)\int_{\tau}^T(\tau-s)^{H-3/2}g(t)dt.
\label{GH}\eaa
 \begin{corollary}\label{corr} Let $\g(t)$ be a process such that $\g(t)$ is $\GG_s$-measurable for all $t$,
and that $\E\int_s^T\g(t)^2dt<+\infty$ for any $T>s$.
Then
\baaa
\int_{s}^T\g(t)dB_H(t)=\int_{s}^T\g(t)dW_H(t)+\int_{s}^T\g(t)\DD R_H(t)dt\\
=\int_{s}^TG_H(\tau,s,T,\g)d B(\tau)+\int_{s}^T\g(t)\DD R_H(t)dt,
\eaaa
and the integrals here converge  in $L_1(\Omega,\GG_{T},\P)$.
\end{corollary}

\par
{\em Proof of Lemma \ref{ThM}} and Corollary \ref{corr} can be found in \cite{D15}.

{\em Proof of Lemma \ref{lemmaI}}. Suppose  that $\g\in\AA_\e(s,T)$, where $\e>$.
Let $\T_\e=\{T_k\}_{k=1}^n$  be the set such as in the definition of $\AA_\e(s,T)$.
By Corollary \ref{corr}, the integrals $\int_{T_{k-1}}^{T_k}\g(t)dB_H(t)$ converge as required for all $k$.
Then the proof follows. $\Box$.

{\em Proof of Lemma \ref{ThE}.}
 Let $\TT$ denotes a finite set of non-random times $\{T_k\}_{k=1}^n\subset [s,T]$, where $n>0$ is an integer, $T_0=s$, $T_n=T$,
and $T_{k+1}\in(T_k,T_k+\e)$; this times are not necessarily equally spaced.
For $\d\in(0,\e)$, let $\Td=\cup_{k=0}^n(T_k,(T_k+\d)\land T)$.   Let $\AA_{\e,\TT,\d}$ be the set of all $\g\in\AA_\e$
such that $\g=0$ for $t\in\Td$.

  Let $I_k= \int_{T_k}^{T_{k+1}}\g_\e(t)dB_H(t)$ and let $\oo I_k=\int_{T_k}^{T_{k+1}}\g_\e(t)dB(t)$.

Let $\T_\e=\{T_k\}_{k=1}^n$  be the set such as in Definition \ref{defS}(ii),   and let
\baaa
I_k=\int_{T_{k-1}}^{T_k}\g(t)dB_H(t)=I_{W,k}+I_{R,k},
\eaaa
where
\baaa
I_{W,k}=\int_{T_{k-1}}^{T_k}\g(t)dW_{H,k}(t),\quad  I_{R,k}=\int_{T_{k-1}}^{T_k}\g(t)dR_{H,k}(t)
\eaaa
and where $W_{H,k}(t)$ and $R_{H,k}$ are defined similarly to $W_H$ and $R_H$ with $[s,T]$ replaced by $[T_{k-1},T_k]$.

Let us prove first the theorem statement for $\g\in \AA_{\e,\d}$.

It suffices to show that
\baa
\E |I_k-\oo I_k|\to 0 \quad\hbox{as}\quad H\to 1/2+0, \quad k=0,1,...,n.
\label{limk}
\eaa
Let us prove (\ref{limk}). Let again $R_{H,k}(t)$ and $\DD R_{H,k}(t)$ be defined similarly to $R_H(t)$ and $\DD R_H(t)$, with the interval $[s,T]$ replaced by the interval $[T_{k-1},T_k]$.
 We have that
$I_k=I_{W,k}+I_{R,k}$, where $I_{W,k}$ and $I_{R,k}$ are defined similarly to the proof of Lemma \ref{ThE}
with the interval $[T_{k-1},T_k]$. Clearly, $\E J_{W,k}=0$.

We have by Lemma \ref{ThM} and by the property of the Riemann--Liouville integral that
$\|\g-G_H(\cdot,T_{k-1},T_k,\g)\|_{L_2(T_{k-1},T_k)}\to 0$ a.s. as $H\to 1/2+0$. In addition,
there exists $c>0$ such that
 \baaa
 \|G_H(\cdot,T_{k-1},T_k,\g)\|_{L_2(T_{k-1},T_k)}\le c\|\g\|_{L_2(T_{k-1},T_k)}\eaaa a.s. for all $H$. Therefore,
 \baaa
 \|\g(t)-G_H(t,T_{k-1},T_k,\g)\|_{L_2(T_{k-1},T_k)}\le 2c\|\g\|_{L_2(T_{k-1},T_k)}\eaaa a.s.  for all $H$.
It follows that
\baaa
\E|I_{W,k} -\oo I_k|^2=\E\int_{T_{k-1}}^{T_k}|\g(t)-G_H(t,T_{k-1},T_k,\g)|^2dt\to 0 \quad\hbox{as}\quad H\to 1/2+0.
\label{II}
\eaaa

For $t\in[T_k\land (T_{k-1}+\d),T_k]$,  we have by Lemma \ref{ThM} that \baaa
\E \DD R_{H,k}(t)^2\le c_H^2\frac{H-1/2}{2}(T_k-T_{k-1}+\d)^{2H-2}=c_H^2\frac{H-1/2}{2}\d^{2H-2}.
\label{DR2}
\eaaa Hence
\baaa
\E|I_{R,k}|\le \left(\E\int_{T_{k-1}}^{T_k}\g(t)^2dt\right)^{1/2} \left(\E\int_{T_k\land (T_{k-1}+\d)}^{T_k}\DD R_H(t)^2dt\right)^{1/2}\to 0\quad\hbox{as}\quad H\to 1/2+0.
\eaaa
Since it holds for all $k$, the theorem statement follows for all $\TT$, $\d$ and $\g_\e\in\A_{\e,\TT,\d}$.
Since any  $\g_\e\in\A_{\e}$ can be represented as $\g_\e=\g^{(1)}+\g^{(2)}$, where $\g^{(k)}\in\A_{\e,\TT_k,\d_k}$, $\k=1,2$,
with an appropriate choice of $\TT_k$ and $\d_k$.
This completes the proof of Lemma \ref{ThE}.
$\Box$

\index{Let us observe that, for any admissible $\g$ such that $\E_{t-\e}\int_{T-\e}^T\g(t)^2dt>0$
the conditional distribution of $X(T)$ given $\F_{T-\e}$ is Gaussian.}

{\em Proof of Theorem \ref{ThNA} (i).} Suppose  that a strategy $\g\in\A_\e$ delivers an arbitrage with the corresponding wealth process $X(t)$ such that $X(0)=0$.

Let  $A_k=\{\int_{T_{k-1}}^{T_k}\g(t)^2dt>0\}$. Let $I_{W,k}$ and $I_{R,k}$ are defined similarly to the proof of Lemma \ref{ThE}
with the interval $[T_{k-1},T_k]$.

  By the definitions, $A_k\in\GG_{T_{k-1}}$. Suppose that
$\P(A_{n}>0)$.

The value $I_{R,n}$ is $\GG_{T_{n-1}}$-measurable, and  the values $\{W_H(t)\}_{t\in [T_{n-1},T]}$ are independent from $\GG_{T_{n-1}}$.
 Since $\g(t)$ is $\GG_{T_{n-1}}$-measurable for $t\in[T_{n-1},T]$ and $\GG_{T_{n-1}}\subset \GG_{T_{n-1}}$, it follows that $I_{W,n}$ and  $I_n$
both have  Gaussian distributions conditionally given $\GG_{T_{n-1}}$. Hence $I_n$ has support on the entire interval $(-\infty,+\infty)$ given $A_n$.

Hence $\P(X(T)<0|A_n)>0$ and $$\P(X(T)<0)=\P(X(T)<0|A_n)\P(A_n)>0.$$ This would be inconsistence with with the supposition that $\g$
delivers an arbitrage. Hence  $\P(A_n)=0$ and  $X(T)=X(T_{n-1})$. Similarly, we obtain that  $\P(A_k)=0$ for all $k$ and $X(T)=0$. This is inconsistence with with the supposition that $\g$
delivers an arbitrage. This completes the proof of Theorem \ref{ThNA}.  $\Box$

{\em Proof of Theorem \ref{ThNA} (ii)} follows from Lemma \ref{ThE}.

\section{Discussion and future developments}
The model presented above  represents a simplest possible model
that allows to illustrate that the arbitrage opportunities vanish for strategies with an arbitrarily small
time delay in information processing.
We leave for future research development of more comprehensive models  and detailed analysis  such as the following.
\begin{enumerate}
\item It could be interesting to investigate if the  discontinuity with respect to $H$ at  $H=1/2$ of stochastic integrals vanishes for piecewise continuous strategies presented in no-arbitrage results obtained in Theorem 4.3 \cite{C} and Theorem 3.21 \cite{B11}.
\item It could be interesting to extend our approach on a more mainstream model with $S(t)=\exp(\mu t+\s B_H(t))$. It is unclear yet how to do this for a setting
with $\F_{t-\e}$-measurable
quantity of shares $\g(t)$ for admissible strategies. However, it is straightforward to consider a model with this prices in a setting with
$\g(t)=\pi(t)/S(t)$, where $\pi(t)$ is $\F_{t-\e}$-adapted.
\end{enumerate}
\subsection*{Acknowledgment} This work  was
supported by ARC grant of Australia DP120100928 to the author.

\end{document}